%% file: D2D_Paper_v5.tex
\documentclass[conference,a4paper]{IEEEtran}

\usepackage{cite}
\usepackage{graphicx,color,epsfig,rotating}
\usepackage{amsfonts,amsmath,amssymb,bbm}
\usepackage{algorithm,algorithmic}
\usepackage{subfigure}
\usepackage{amsmath}
\usepackage{cite}
\usepackage{mdwtab}
\usepackage{placeins}
\usepackage{psfrag, graphicx}
\usepackage[latin1]{inputenc}
\usepackage{amssymb}
\usepackage{multirow}

\input{macros}

\newtheorem{defn}{Definition}
\newtheorem{theorem}{Theorem}

\begin{document}

\sloppy

\title{Optimal Throughput-Outage Trade-off in Wireless One-Hop Caching Networks}

\author{
  \IEEEauthorblockN{Mingyue Ji}
  \IEEEauthorblockA{Department of Electrical Engineering\\
    University of Southern California\\
    Email: mingyuej@usc.edu} 
  \and
  \IEEEauthorblockN{Giuseppe Caire}
  \IEEEauthorblockA{Department of Electrical Engineering\\
    University of Southern California\\
    Email: caire@usc.edu} 
  \and
  \IEEEauthorblockN{Andreas F. Molisch}
  \IEEEauthorblockA{Department of Electrical Engineering\\
    University of Southern California\\
    Email: molisch@usc.edu} 
}



\maketitle

\begin{abstract}
We consider a wireless device-to-device (D2D) network where the nodes have cached information from a library of possible files. 
Inspired by the current trend in the standardization of the D2D mode for 4th generation wireless networks, we restrict to one-hop communication:
each node places a request to a file in the library, and downloads from some other node which has the requested file in its cache through a direct communication link, 
without going through a base station. We describe the physical layer communication through a simple ``protocol-model'', based on interference avoidance (independent set scheduling). For this network we define the outage-throughput tradeoff problem and characterize the optimal scaling laws for various regimes where both the number 
of nodes and the files in the library grow to infinity.  
\end{abstract}

\section{Introduction}
\label{section: intro}

Wireless data traffic is increasing dramatically, with a 6600\% increase predicted for the next five years. This is mainly due to wireless video streaming. 
Traditional methods for increasing the area spectral efficiency, such as use of more spectrum and increase in the number of base stations, 
are either insufficient to provide a suitable capacity increase, or are too expensive. 
There is thus a great need to explore alternative transmission strategies. 

While live streaming is a negligible portion of the wireless video traffic, the bulk is represented by asynchronous {\em video on demand}, where users request video files from some
library (e.g., the top 100 titles in Netflix or Amazon Prime) at arbitrary times. Therefore, trivial uncoded multi-casting (i.e., serving many users with a single downlink transmission)
cannot be exploited in this context. 
One of the most promising approaches is {\em caching}, i.e., storing popular content at, or close to, the users. 
As has been pointed out, in \cite{CommMag}, 
caching can be used in lieu of backhaul for providing content to users; for example, messages (e.g., video files) can be delivered during off-peak hours to the caches while the files can be used during peak traffic hours. 
In this paper we will particularly concentrate on caching at mobile devices, 
which is enabled by the availability of tens and even hundreds of GByte of largely under-utilized storage space in smartphones, 
tablets, and laptops. 

Recently, a coded multicasting scheme exploiting caching at the user nodes was proposed in \cite{maddah2012fundamental}. In this scheme, a combination of caching and coded multicast transmission from a single base station 
is used in order to satisfy all users requests at the same time. The construction of the caches is combinatorial, and changing even a finale file in the library requires a complete reconfiguration of the user caches. Therefore, the approach is not yet practical. 
In this paper we focus on a quite different alternative that involves random independent caching at the user nodes and device-to-device (D2D) communication. We restrict 
to one-hop communication, inspired by the current trend in the standardization of a D2D mode for 4th generation cellular systems \cite{wu2010flashlinq}. 

A relevant and related work is given in \cite{gitzenis2012asymptotic}, 
where multi-hop D2D communication is considered under a distance-based protocol transmission model \cite{gupta2000capacity}. 
If the aggregate distributed storage space in the network is larger than the total size of 
all messages, then it can be guaranteed that all users can be served by this network. 
Under assumption of a Zipf request distribution with parameter $\gamma_r$ (to be defined later), the author of \cite{gitzenis2012asymptotic} 
design a deterministic duplication caching scheme and a multi-hop routing scheme that achieves order-optimal 
average throughput. 

Since we consider only single-hop communication, requiring that all users are actually served for any request is too constraining. Therefore, we
generalize the problem by introducing the possibility of outages, i.e., that some request is not served. 
For the system defined in Section \ref{section: network model} we define the 
outage-throughput region and obtain achievable scaling laws and upper bounds which are tight enough to characterize the constant of the leading term. 
Simulations agree very well with the scaling law leading constants. We also compare the D2D system under investigation with the performance of the 
coded multicast of \cite{maddah2012fundamental} 
and with naive broadcasting from the cellular base station (independent messages), which can be regarded as today's 
state of the art.~\footnote{Notation: given two functions $f$ and $g$, we say that: 1) 
$f(n) = O\left(g(n)\right)$ if there exists a constant $c$ and integer $N$ such that 
$f(n)\leq cg(n)$ for $n>N$. 2) $f(n)=o\left(g(n)\right)$ if $\lim_{n \rightarrow \infty}\frac{f(n)}{g(n)} = 0$. 
3) $f(n) = \Omega\left(g(n)\right)$ if $g(n) = O\left(f(n)\right)$. 4) 
$f(n) = \omega\left(g(n)\right)$ if $g(n) = o\left(f(n)\right)$. 
5) $f(n) = \Theta\left(g(n)\right)$ if $f(n) = O\left(g(n)\right)$ and $g(n) = O\left(f(n)\right)$.}

A similar setting was investigated by  \cite{golrezaei2012wireless}, where only the sum throughput was 
considered irrespectively of user outage probability. Furthermore, in \cite{golrezaei2012wireless}  a heuristic random caching policy according to another 
Zipf distribution with a possibly different parameter $\gamma_c$ was considered. The results showed that the optimal throughput occurred 
when $\gamma_r \neq \gamma_c$, but the throughput order by this heuristic random caching policy is generally suboptimal. 
More importantly, the total sum throughput is not a sufficient characterization of the performance of a one-hop D2D caching network: 
in certain regimes of the number of users and file library size it can be shown that to achieve a high throughput only a small portion of the users should 
be served while leaving the majority of the users are in outage. In contrast, our outage-throughput tradeoff region is able to capture the notion of fairness, since it focuses
on the minimum per-user average throughput. 

The paper is organized as follows. Section~\ref{section: network model} introduces the network model and the precise problem formulation of the throughput-outage trade-off in wireless D2D networks. Section~\ref{section: Achievable Throughput-Outage Trade-off} presents the achievable throughput-outage trade-off. The outer bound of this trade-off is discussed in Section~\ref{section: The Converse of Throughput-Outage Trade-off}. We discuss our reuslts in Section~\ref{sec: Discussion and Conclusion}. 


\section{Network Model and Problem Formulation}
\label{section: network model}

\begin{figure}
\centering
\subfigure[]{
\centering \includegraphics[width=3.5cm]{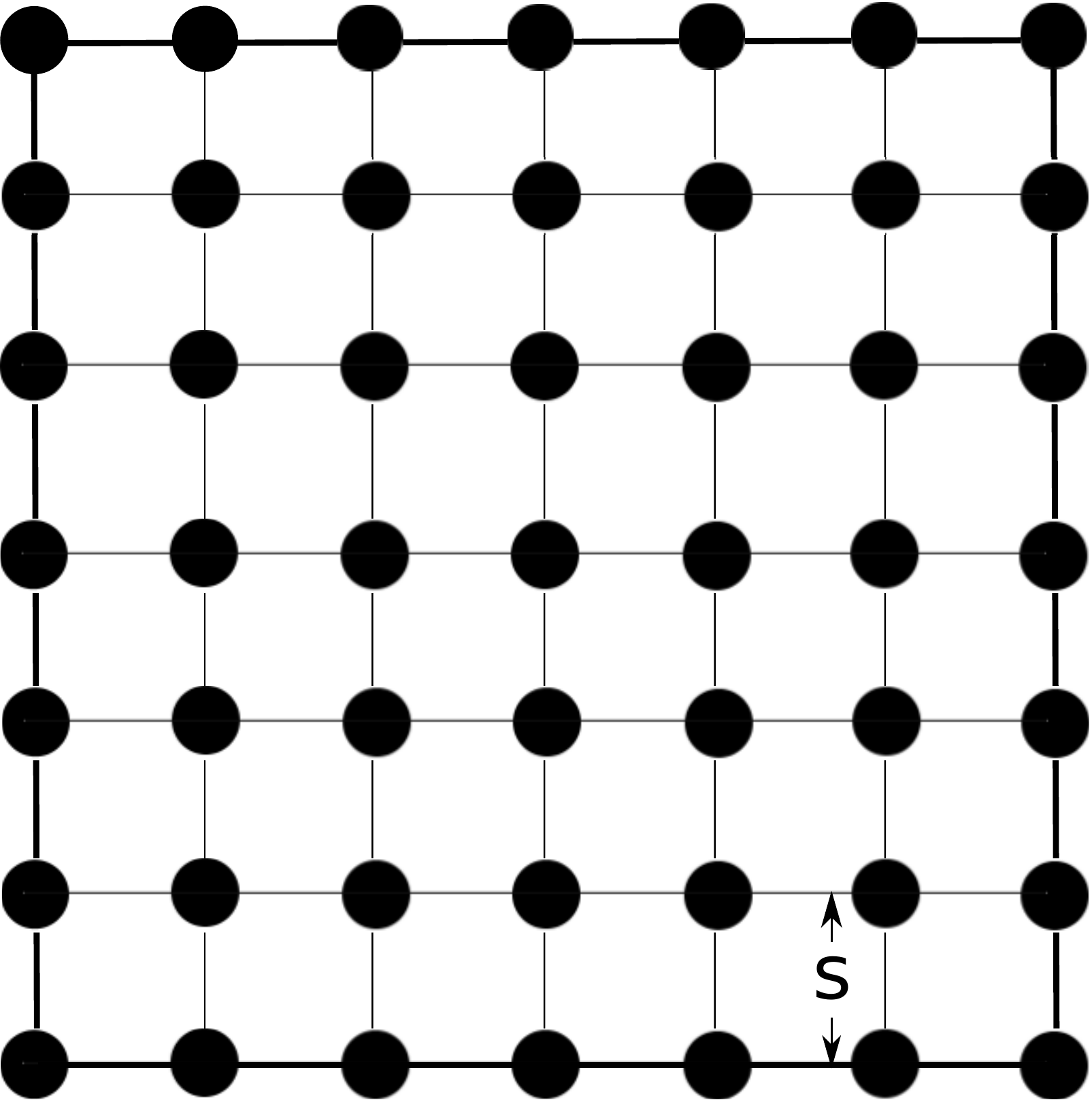}
\label{fig: Grid_Network_D2D}
}
\subfigure[]{
\centering \includegraphics[width=3.5cm]{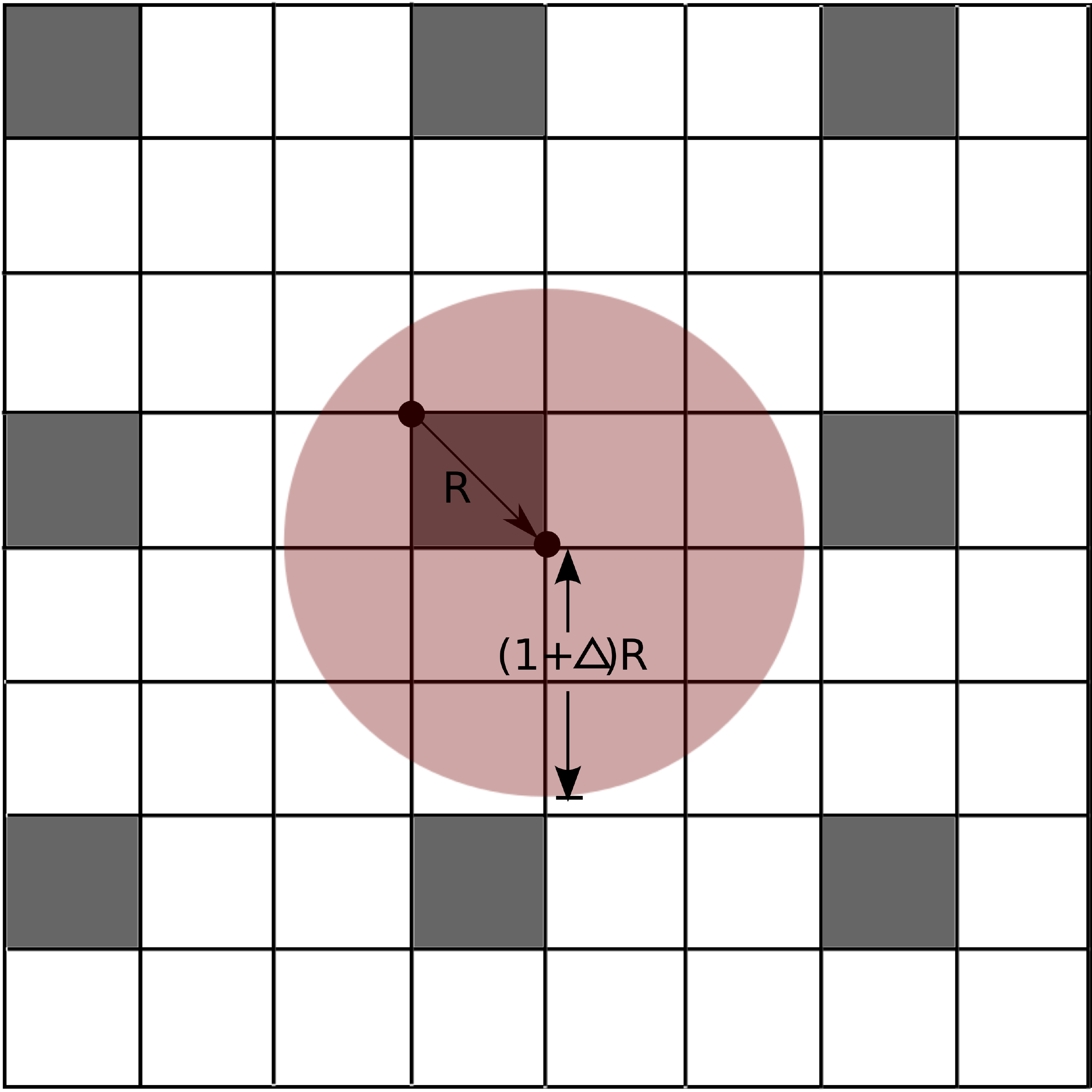}
\label{fig: Grid_TDMA}
}
\caption{a)~Grid network with $n=49$ nodes (black circles) with minimum separation $s = \frac{1}{\sqrt{n}}$. 
b)~An example of single-cell layout and the interference avoidance TDMA scheme. 
In this figure, each square represents a cluster. 
The gray squares represent the concurrent transmitting clusters. 
The red area is the disk where the protocol model imposes no other concurrent transmission. 
$R$ is the worst case transmission range and $\Delta$ is the interference parameter. 
We assume a common $R$ for all the transmitter-receiver pairs. 
In this particular example, the TDMA parameter is $K=9$.}
\end{figure}

\begin{figure}
\centering
\includegraphics[width=8cm]{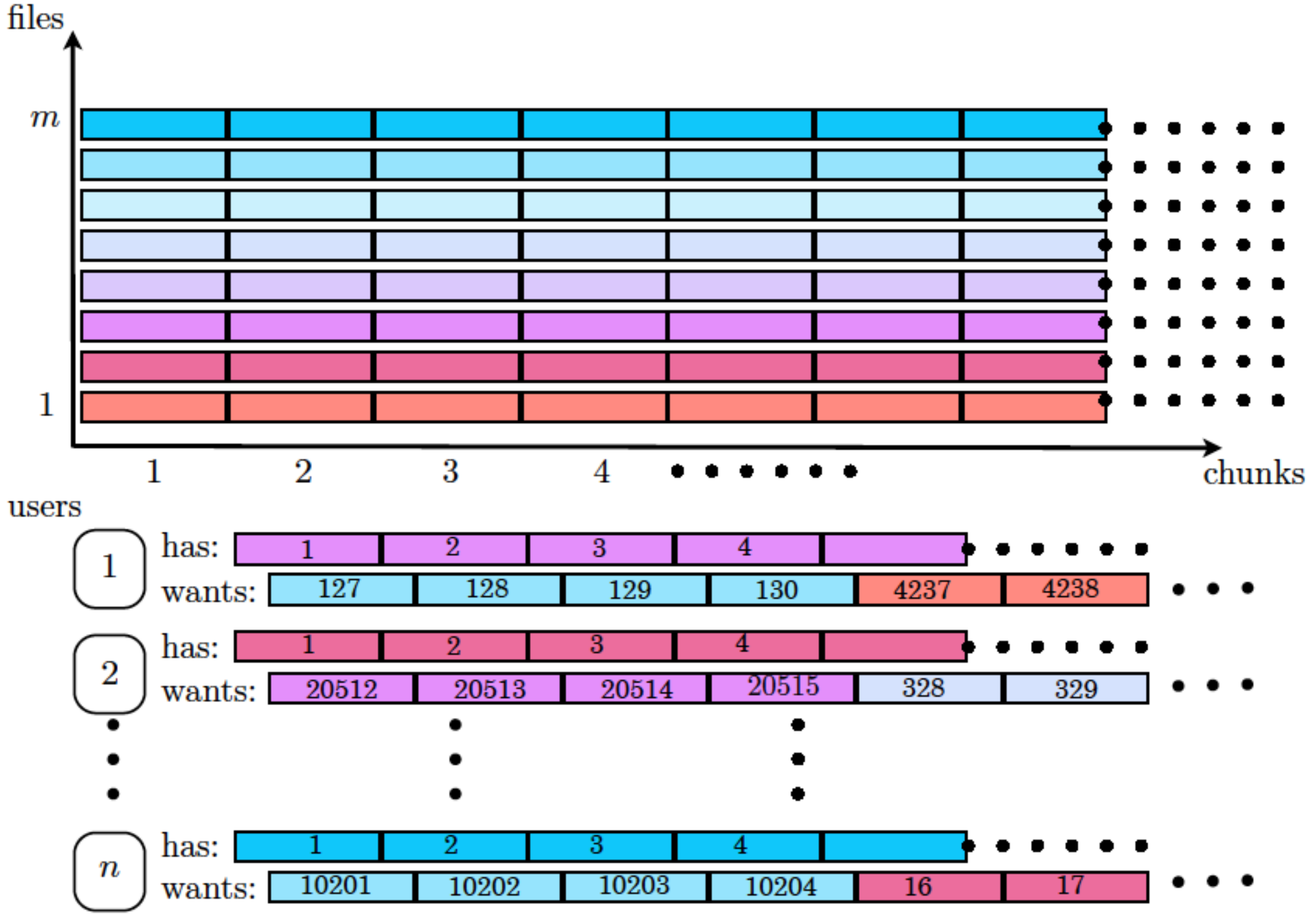} 
\caption{Qualitative representation of our system assumptions: each user caches an entire file, formed by an arbitrarily large number of chunks.
Then, users place random requests of finite sequences of chunks from files of the library, or random duration and random initial points.}
\label{library-and-asynchronous-demands}
\end{figure}

We consider a dense network deployed over a unit-area square and formed by $n$ nodes $\Uc = \{1, \ldots, n\}$ 
placed on a regular grid with minimum node distance  
$1/\sqrt{n}$ (see Fig.~\ref{fig: Grid_Network_D2D}).
Each user $u \in \Uc$ makes a request to a file $f_u \in \Fc = \{1, \ldots, m\}$ in an i.i.d. manner, 
according to a given request probability mass function $P_r(f)$. 
In order to model the asynchronism of video on demand 
and forbid any form of ``for-free'' multicasting gain by ``overhearing'' transmissions dedicated to others, 
we assume that each file in the library is formed by $L$ ``chunks''. For example, in current video streaming protocols such as DASH 
\cite{wu2010flashlinq}, the video file is split into segments which are sequentially downloaded by the streaming users. 
The chunk downloading time is equal to the chunk playback time, but chunks may correspond to different bit-rates, 
depending on the video coding quality. Then, we assume that requests are strongly asynchronous: each user downloads a segment of length $L'$ of a long file
of $L$ chunks. We measure the cache size in files, and let first $L \rightarrow \infty$ and then study the system scaling laws for 
$n,m \rightarrow \infty$, with fixed $L' \leq \infty$. 
Hence, the probability of useful overhearing vanishes, while the probability that two users request the same {\em file} 
depends on the library size $m$ and on the request distribution $P_r$. In short, this is a conceptual way to decouple the overlap of the demands 
with the overlap of concurrent transmissions, which would be difficult if not impossible to exploit in a practical system.   
For the sake of simplicity, we assume that the user caches contain $M = 1$ files ($ML$ chunks) in the analysis. 
Fig.~\ref{library-and-asynchronous-demands} shows qualitatively our model assumptions. 

\begin{defn} {\bf (Protocol model)}
\label{definition ProtocolModel}
If a node $i$ transmits a packet to node $j$, then the transmission is successful if and only if
\begin{itemize}
\item The distance between $i$ and $j$ is less than $R$.
\be
	d(i,j) \leq R.
\ee
\item For any other node $k$ that is transmitting simultaneously,
\be
	d(k,j) \geq (1+\Delta) R.
\ee
\end{itemize}
$R$ is the transmission range and $\Delta > 0$ is an interference control parameter. 
Nodes send data at a constant rate of $C$ bit/s/Hz a successful transmission.
\hfill $\lozenge$
\end{defn}

In our model we do not consider power control (which would allow different transmit powers, and thus transmission ranges), for each user. Rather, we treat $R$ as a design parameter that can be set as a function of $m$ and $n$, but which cannot vary between users.  
 
\begin{defn} {\bf (Network)} A network if formed by a set of user nodes $\Uc$, a set of helper nodes
$\Hc = \{1, \ldots, r\}$ and a set of files $\Fc = \{1, \ldots, m\}$. Nodes in $\Uc$ and $\Hc$ are placed in a two-dimensional unit-square region, 
and their transmissions obeys the protocol model. Helper nodes are only transmitters, user nodes can be transmitters and receivers. 
In general, all $n (n-1)$ directed links between all user nodes and all $r n$ directed links between the helper nodes and the user nodes, 
together with the protocol model define a interference (conflict) graph. 
Only the links in an independent set in the interference graph can be 
active simultaneously.
\hfill $\lozenge$
\end{defn}
 
\begin{defn}
{\bf (Cache placement)} The cache placement $\Pi_c$ is a rule to assign files from the library $\Fc$ to the user nodes 
$\Uc$ and the helper nodes $\Hc$ with ``replacement'' (i.e., with possible replication). 
Let $\Gsf = \{\Uc \cup \Hc, \Fc, \Ec\}$ be a bipartite graph with 
``left'' nodes $\Uc \cup \Hc$, ``right'' nodes $\Fc$ and edges $\Ec$ such that $(u,f) \in \Ec$ indicates that file 
$f$ is assigned to the cache of user node $u$ and $(h,f) \in \Ec$ indicates that file $f$ is assigned to the 
cache of helper node $h$. A bi-partite cache placement graph $\Gsf$ is feasible if the degree of each left node (user or helper) 
is not larger than its maximum cache capacity $M$. Let $\Gc$ denote the set of all feasible bi-partite graphs $\Gsf$. 
Then,  $\Pi_c$ is a probability mass function over $\Gc$,  i.e., a particular cache placement $\Gsf \in \Gc$ is assigned   with probability $\Pi_c(\Gsf)$. 
\hfill $\lozenge$
\end{defn}
  
Notice that deterministic cache placements are special cases, corresponding to 
deterministic probability mass functions, a single probability mass equal to 1 on the desired $\Gsf$. 
In contrast, we will be interested in ``decentralized'' random caching placements with no helpers 
constructed as follows: each user node $u$ selects its cache content in an i.i.d. manner, by independently generating 
$M=1$ random file indices with the same caching probability mass function $\{P_c(f) : f \in \Fc \}$. 

\begin{defn} {\bf (Random requests)} At each request time (integer multiples of some fixed (large) integer $L'$), 
each user $u \in \Uc$ makes a request to a segment of length $L'$ of chunks from file $f_u \in \Fc$, 
selected independently with probability $P_r$. The set of current requests $\fsf = (f_{1}, \ldots, f_{n})$ is 
therefore a random vector taking on values in $\Fc^n$, with product joint probability mass function 
$\PP(\fsf = (f_{1}, \ldots, f_{n})) = \prod_{i=1}^n P_r(f_{i})$.
\hfill $\lozenge$
\end{defn} 

In this paper, we assume $P_r(f)$ follows a Zipf distribution with 
parameter $0 < \gamma_r < 1$, i.e., any node requests file $f$ with probability  
$\frac{f^{-\gamma_r}}{H(\gamma_r,1,m)}$,  where we define $H(\gamma,a,b) = \sum_{f=a}^b \frac{1}{i^\gamma}$ and $f=1, \cdots, m$. 

\begin{defn} 
{\bf (Transmission policy)} The transmission policy $\Pi_t$ is a rule to activate the D2D links in the network.
Let $\Lc$ denote the set of all directed links.  Let $\Ac \subseteq 2^\Lc$ the set of all possible feasible subsets of links 
(this is a subset of the power set of $\Lc$, formed by all sets of links corresponding to independent sets in the 
network interference graph).  Let $\Asf \subset \Ac$ denote a feasible set of simultaneously active links according to the protocol model. 
Then, $\Pi_t$ is a conditional probability mass function over $\Ac$ given $\fsf$ (requests) and 
$\Gsf$ (cache placement),  assigning probability $\Pi_t(\Asf |\fsf, \Gsf)$ to $\Asf \in \Ac$. \hfill $\lozenge$
\end{defn}

We may think of $\Pi_t$ as a way of scheduling simultaneously compatible sets of links 
(subject to the protocol model).  The scheduling slot duration is generally much shorter than the chunk playback duration. 
Invoking a time-scale decomposition, and provided that enough buffering is used at the receiving end, 
we can always match the average throughput (expressed in information bit/s) per user with the average
source coding rate at which the video file can be streamed to a given user. 
Hence, while the chunk delivery time is fixed (e.g.,  one chunk per 0.5 seconds) 
the ``quality''  at which the video is streamed and reproduced at the user end 
depends on the user average throughput.  Therefore, in this scenario we are concerned with the ergodic (i.e., {\em long-term average}) 
throughput  per user.

\begin{defn}
\label{definition: throughput} 
{\bf (Useful received bits per slot)} For given $P_r$, $\Pi_c$ and $\Pi_t$, and user $u  \in \Uc$ we define 
the random variable $T_u$ as the number of useful  received information bits per slot unit time 
by user $u$ at a given scheduling time (irrelevant because of 
stationarity). This is given by 
\begin{equation}  \label{useful-throughput-i}
T_u = \sum_{v : (u,v) \in \Asf} c_{u,v} 1\{ f_u \in  \Gsf(v) \}
\end{equation}
where $f_u$ denotes the file requested by user node $u$,  $c_{u,v}$ denotes the rate of the link $(u,v)$, 
and $\Gsf(v)$ denotes the content of the cache of node $v$, i.e., the neighborhood of left node $v$ in  the cache placement graph $\Gsf$. 
\hfill $\lozenge$
\end{defn}

Consistently with the protocol model, $c_{u,v}$ depends only on the active link $(u,v) \in \Asf$ and not on the whole set of active 
links $\Asf$. Furthermore, we shall obtain most of our results under the simplifying assumption 
(usually made under the protocol model) that $c_{u,v} = C$ for all $(u,v) \in \Asf$. 
The indicator function $1\{ f_u \in \Gsf(v) \}$ expresses the fact that only the bits relative to the file 
$f_u$ requested by user $u$ are ``useful'' and count towards the throughput. 
It is obvious that scheduling links $(u,v)$ for which $f_u \notin \Gsf(v)$  
is useless for the sake of the throughput defined as above. Hence, we could restrict our transmission policies
to those activating only links $(u,v)$ for which $f_u \in \Gsf(v)$. These links are referred to as 
``potential links'', i.e., links potentially carrying useful data. Potential links included in $\Asf$ are ``active links'', 
at the given scheduling slot. 

The average throughput for user node $u \in \Uc$ is given by $\overline{T}_u = \EE[T_u]$, where expectation  is with respect to the random triple 
$(\fsf, \Gsf, \Asf) \sim \prod_{u=1}^n P_r(f_u) \Pi_c(\Gsf) \Pi_t(\Asf|\fsf,\Gsf)$. Next, we define the condition of ``user in outage'' consistently with the 
qualitative  system description given before. In particular, consider a user $u$ and its useful received bits per slot $T_u$.
We say that user $u$ is in outage if $\EE[T_u | \fsf, \Gsf] = 0$. This condition captures the event that no link $(u,v)$ with $f_u \in \Gsf(v)$ is scheduled 
with positive probability, for given set of requests $\fsf$ and cache placement $\Gsf$. In other words, a user $u$ for which $\EE[T_u | \fsf, \Gsf] = 0$ 
experiences  a ``long'' lack of service (zero rate), as far as the cache placement is $\Gsf$ and the request vector is $\fsf$. 

\begin{defn}
{\bf (Number of nodes in outage)}  The number of  nodes in outage is given by 
\begin{equation} \label{outage-nodes}
N_o = \sum_{u \in \Uc} 1\{ \EE[T_u |\fsf, \Gsf]  = 0\}. 
\end{equation}
Notice that $N_o$ is a random variable, function of $\fsf$ and $\Gsf$. 
\hfill $\lozenge$
\end{defn}

\begin{defn}
\label{definition: outage}
{\bf (Average outage probability)}  The average (across the users) outage probability is given by 
\begin{equation} \label{outage-probability}
p_o = \frac{1}{n} \EE[ N_o ] = \frac{1}{n} \sum_{u \in \Uc} \PP\left ( \EE[T_u |\fsf, \Gsf]  = 0 \right ). 
\end{equation}
\hfill $\lozenge$
\end{defn}

Here, we focus on max-min fairness, i.e., we express the outage-throughput tradeoff in terms of the minimum 
average user throughput, defined as
\begin{equation}
\overline{T}_{\rm min} = \min_{u \in \Uc} \; \left \{ \overline{T}_u  \right \}. 
\end{equation}
At this point we can define the performance tradeoffs that we wish to characterize in this work: 

\begin{defn} {\bf (Outage -- Throughput Tradeoff)}  
For a given network and request probability distribution $P_r$, 
an outage-throughput pair $(p,t)$ is {\em achievable} if there exists a cache placement $\Pi_c$ and a transmission policy $\Pi_t$ with outage probability
$p_o \leq p$ and minimum per-user average throughput $\overline{T}_{\min} \geq t$. The outage-throughput achievable region
$\Tc(P_r,n,m)$ is the closure of all achievable outage-throughput pairs $(p,t)$. 
In particular, we let $T^*(p) = \sup \{ t : (p, t) \in \Tc(P_r,n,m) \}$. 
\hfill $\lozenge$
\end{defn}

Notice that $T^*(p)$ is the result of the following optimization problem:
\begin{eqnarray} \label{sucaminchia}
\mbox{maximize} & & \overline{T}_{\min} \nonumber \\
\mbox{subject to} & & p_o \leq p, 
\end{eqnarray}
where the maximization is with respect to the cache placement and transmission policies $\Pi_c, \Pi_t$. 
Hence, it is immediate to see that $T^*(p)$ is non-decreasing in the range of feasible outage probability, 
which in general is the interval $[p_{o, \min}, 1]$ for some $p_{o,\min} \geq 0$. 
Whether $p_{o,\min}$ is equal to 0 or it is strictly positive depends on the model assumptions. 
We say that that an achievable point $(p,t)$ dominates an achievable point $(p',t')$  if $p\leq p'$ and $t \geq t'$ where at least one 
of the inequalities is strict.   As usual, the Pareto boundary of $\Tc(P_r,n,m)$ consists of all achievable points that are 
not dominated by other achievable points. 

\section{Achievable Outage-Throughput Trade-off}
\label{section: Achievable Throughput-Outage Trade-off}

We obtain an inner lower bound on the achievable throughput-outage tradeoff by considering specific transmission policy based on 
clustering and independent random caching.

{\bf Clustering:}  the network is divided into clusters of equal size, denoted by $g_c(m)$ and 
independent of the users' demands and  cache placement realizations.
A user can only look for the requested file inside the corresponding cluster.  
If a user can find the requested file inside the cluster, we say there is one \emph{potential link} in this cluster.  
Moreover, if a cluster contains at least one potential link, we say that this cluster is \emph{good}. 
We use an {\em interference avoidance} scheme for which at most one transmission is allowed in each cluster, 
on any time-frequency slot (transmission resource). 
Potential links inside the same cluster are scheduled with equal probability (or, equivalently, in round robin), such that all users have the samel throughput $\overline{T}_u = \overline{T}_{\min}$. To avoid interference between clusters, we use a time-frequency reuse scheme \cite[Ch. 17]{molisch2011wireless} with parameter $K$ 
as shown in Fig.~\ref{fig: Grid_TDMA}. In particular, we can pick $K = \left(\left\lceil\sqrt{2}(1+\Delta)\right\rceil+1\right)^2$. 

{\bf Random Caching:} each node randomly and independently caches one file according to a common probability distribution function $P_c$. We shall find the optimal $P_c$
that maximizes the achievable $\overline{T}_{\min}$ under the clustering scheme.

In the rest of this paper, unless said otherwise,  is assumed that $n,m \rightarrow \infty$ in some way (to be specified later). 
Proofs are omitted for the sake of space limitation, and are provided in \cite{mingyue2013}.
We start by  characterizing the optimal random caching distribution under the clustering transmission scheme.

\begin{theorem}
\label{theorem: optimal caching distribution}
Under the model assumptions and the clustering scheme, the optimal caching distribution $P_c^*$ that maximize the probability $p_u^c$ 
that any user $u \in \Uc$ finds its requested file inside its corresponding cluster is given by
\begin{equation}
P_c^*(f) = \left[1 - \frac{\nu}{z_{f}}\right]^+,  \;\;\; f = 1,\ldots, m,
\end{equation}
where $\nu = \frac{m^*-1}{\sum_{j=1}^{m^*} \frac{1}{z_{j}}}$, 
$z_{j} = P_r(j)^{\frac{1}{g_c(m) - 2}}$, 
and $m^* = \Theta \left (\min \{\frac{1}{\gamma_r}g_c(m), m\}\right )$.
\hfill  $\square$
\end{theorem}

Next, we distinguish the different regimes of small library size, large library size and very large library size. Letting $m$ vary as a function of $n$, and $\xi$ indicate some strictly positive constant, we have
\begin{eqnarray}
\lim_{n \rightarrow \infty} \frac{m}{n^\alpha} & = & 0 , \;\; \mbox{small library} \label{small-lib} \\
0 < \xi  \leq \lim_{n \rightarrow \infty} \frac{m}{n^\alpha}  & \leq & \left ( \frac{\gamma_r^{\gamma_r}}{1 - \gamma_r} \right )^\frac{\alpha}{2-\gamma_r}, \;\; \mbox{large library} \\
\lim_{n \rightarrow \infty} \frac{m}{n^\alpha}  & > & \left ( \frac{\gamma_r^{\gamma_r}}{1 - \gamma_r} \right )^\frac{\alpha}{2-\gamma_r}, \;\; \mbox{very large library} \nonumber \\
& &
\end{eqnarray}
where we define $\alpha = \frac{2- \gamma_r}{1 - \gamma_r}$. Then, we have:

\begin{theorem} \label{theorem: 1}
In the small library regime, the achievable outage-throughput trade-off achievable by random caching and the clustering scheme behaves as:
\begin{align}
\label{eq: theorem 1}
&T^*(p) \geq \notag\\
& \left\{\begin{array}{ll}
\frac{C}{K}\frac{1}{\rho_1 m} +  \delta_1(m), & \;\; p = (1-\gamma_r) e^{\gamma_r - \rho_1} \\
\frac{CA}{K} \frac{1}{m (1-p)^{\frac{1}{1-\gamma_r}}} + \delta_2(m),  & \;\; p = 1 - {\gamma_r}^{\gamma_r} \left(\frac{g_c(m)}{m}\right)^{1-\gamma_r}, \\
\frac{CB}{K } m^{-1/\alpha} + \delta_3(m),  & \;\; 1 - {\gamma_r}^{\gamma_r} \rho_2^{1-\gamma_r} m^{-1/\alpha} \leq  \\
& p \leq 1 - a(\gamma_r)  m^{-1/\alpha},  \\
\frac{CD}{K} m^{-1/\alpha} + \delta_4(m),  & \;\; p \geq 1 - a(\gamma_r) m^{-1/\alpha}
\end{array}\right.
\end{align}
where $a(\gamma_r) = 
{\gamma_r}^{\gamma_r}\left(\frac{1-\gamma_r}{{\gamma_r}^{\gamma_r}}\right)^{1/\alpha}$, 
$A = {\gamma_r}^{\frac{\gamma_r}{1-\gamma_r}}$,
$B = \frac{{\gamma_r}^{\gamma_r}\rho_2^{1-\gamma_r}}{1+{\gamma_r}^{\gamma_r}\rho_2^{2-\gamma_r}}$, 
$D =  \frac{a(\gamma_r)}{1+a(\gamma_r)\left(\frac{1-\gamma_r}{\gamma_r^{\gamma_r}}\right)^{\frac{1}{2-\gamma_r}}}$ and where $\rho_1$ and $\rho_2$ are positive parameters satisfying
$\rho_1 \geq \gamma_r$ and $\rho_2 \geq  \left(\frac{1-\gamma_r}{\gamma_r^{\gamma_r}}\right)^{\frac{1}{2-\gamma_r}}$. 
The cluster size $g_c(m)$ is any function of $m$ satisfying 
$g_c(m) = \omega\left(m^{1/\alpha} \right)$ and $g_c(m) \leq \gamma_r m$. The functions $\delta_i(m)$ $i = 1,2,3,4$ are vanishing for $m \rightarrow \infty$ with the following orders $\delta_1(m) = o(1/m)$, $\delta_2(m) = o\left(\frac{1}{m(1-p)^{\frac{1}{1-\gamma_r}}} \right)$,
$\delta_3(m)$, 
$\delta_4(m) = o\left(m^{-1/\alpha}\right)$.
\hfill $\square$
\end{theorem}

The results for the large and very large library regimes can be found in \cite{mingyue2013}.

\section{Outer Bound}
\label{section: The Converse of Throughput-Outage Trade-off}

Under the assumptions of protocol model (see Definition~\ref{definition ProtocolModel}) 
and one-hop transmission, we can provide an outer bound on the outage-throughput tradeoff
$(p, T^{\rm ub}(p))$ such that the ensemble of such points for $p \in [0,1]$ dominates the optimal trade-off, i.e., the ensemble of solutions of (\ref{sucaminchia}). 
We have:

\begin{theorem}
\label{theorem: 2}
In the small library regime, the set of points defined below dominates the 
optimal throughput-outage tradeoff:
\begin{align}
&T^{\rm ub}(p) = \notag\\
&\left\{\begin{array}{ll} 
\frac{16C}{\Delta^2m(1-p)^{\frac{1}{1-\gamma_r}}} + \delta_5(m) , & \;\; p = 1 - \left(\frac{g_R(m)}{n}\right)^{1-\gamma_r}, \\
\min\left\{ \frac{16C}{\Delta^2 m (1-p)^{\frac{1}{1-\gamma_r}}},  \right. & \\
\left. f_1(\rho_3) m^{-1/\alpha} \right\} + \delta_6(m), & 1 - {\rho_3}^{1-\gamma_r} m^{-1/\alpha} \leq \\ 
& p < 1 - {\rho_4}^{1-\gamma_r} m^{-1/\alpha}, \\
f_1(\rho_4) m^{-1/\alpha} + \delta_7(m) , & 1 - {\rho_4}^{1-\gamma_r} m^{-1/\alpha} \leq p \leq 1, \\
\end{array} \right.
\end{align}
where $\rho_3$ is a positive parameter and $\rho_4$ is the solution of the equation 
\begin{align}
&\left(\left(1+\frac{3\Delta}{2}\right)^2 \rho \right)^{2-\gamma_r} \notag\\
&= \log\left(1+(2-\gamma_r)\left(\left(1+\frac{3\Delta}{2}\right)^2 \rho \right)^{2-\gamma_r}\right),
\end{align}
with respect to $\rho$,  $g_R(m)$ is any function such that $g_R(m) = \omega\left(m^{1/\alpha} \right)$ and $g_R(m) \leq \frac{16}{\Delta^2} n$, 
$f_1(\rho) = \frac{16C}{\Delta^2 \rho} \left(1-\exp\left(-\left(1+\frac{3\Delta}{2}\right)^{2(2-\gamma_r)}{\rho}^{2-\gamma_r}\right)\right)$, and 
$\delta_5(m) = o\left(\frac{1}{m(1-p)^{\frac{1}{1-\gamma_r}}} \right)$,
$\delta_6(m)$, 
$\delta_7(m) = o\left(m^{-1/\alpha} \right)$.  
\hfill  $\square$
\end{theorem}

The results of other regimes of $m$ can be found in \cite{mingyue2013}. In all cases, 
notice that the scaling laws of the throughput and outage probability with respect to $m \rightarrow \infty$ coincide and are therefore tight up to
some gap in the constants of the leading terms. 

\section{Discussion}
\label{sec: Discussion and Conclusion}

\begin{figure}[ht]
\centerline{\includegraphics[width=8cm]{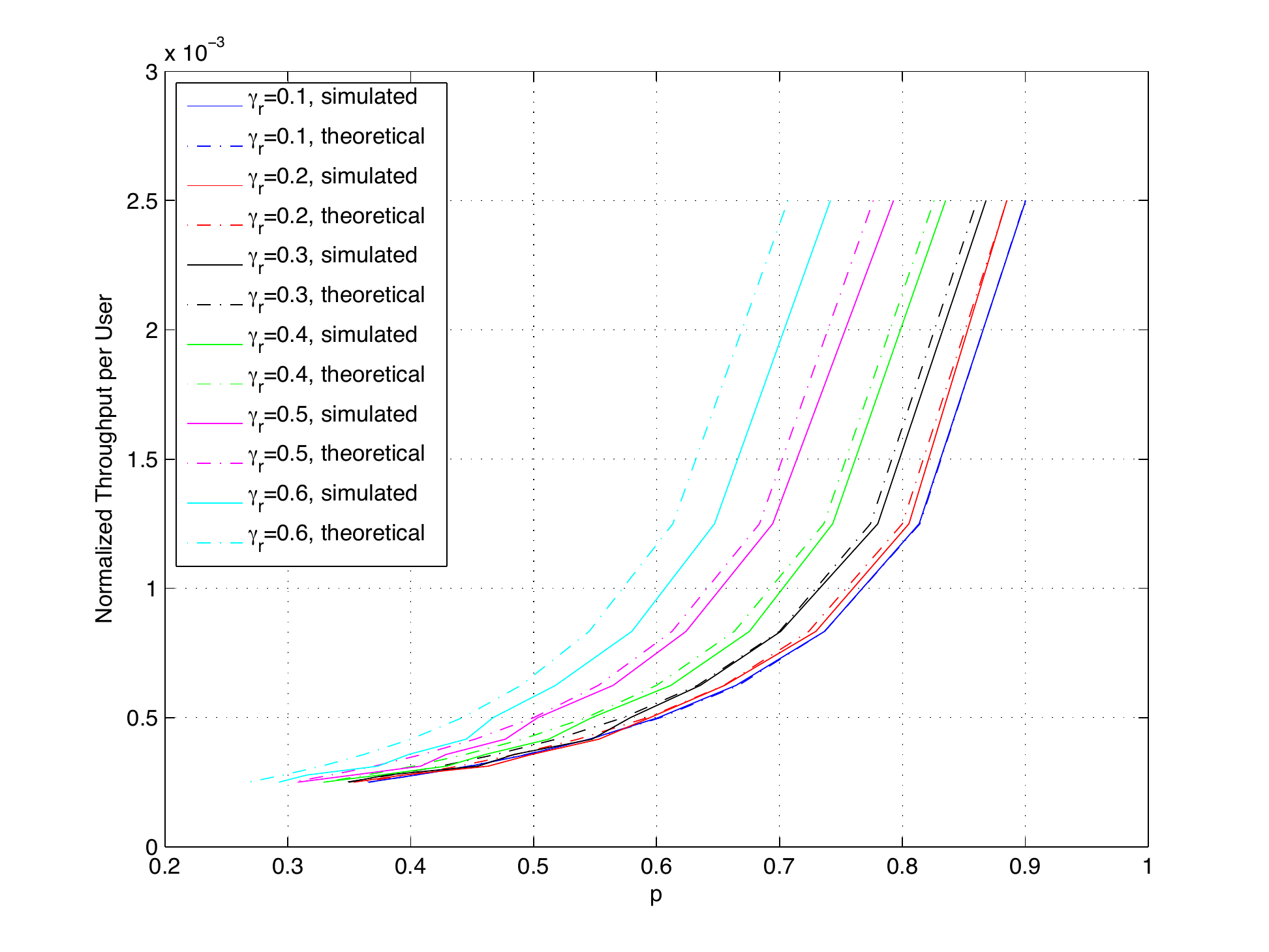}}
\caption{In this figure, we show a comparison between the normalized theoretical result and normalized simulated result in terms of the minimum throughput per user v.s. outage probability constraint. The normalization is by $C$. We assume $m=1000$, $n=10000$, $K = 4$. 
The parameter $\gamma_r$ for the Zipf distribution varies from $0.1$ to $0.6$. The theoretical curve is the plot of the dominate term in (\ref{eq: theorem 1}) normalized by $C$.}
\label{fig: result}
\end{figure}

In this section, we focus on the regime of small library as provided in \emph{Theorem}~\ref{theorem: 1}. Specifically, we consider the regime of constant outage probability constraint ($0 < p < 1$ and $g_c(m) \propto m$). We realistically assume that $m=1000$ and $n=10000$ (this corresponds to one node every $10 \times 10$m, in a $1$ km$^2$ area). 
Moreover, we let $K = 4$. 
The simulation of the normalized throughput per user is shown in Fig.~\ref{fig: result}. This simulation shows that even for practical $m$ and $n$, the dominate term in (\ref{eq: theorem 1}) accurately captures the system behavior.

It is clear that the naive broadcasting from the cellular base station gives a minimum per user throughput at $\Theta\left(\frac{1}{n}\right)$ without outage. In \cite{maddah2012fundamental}, where the authors assume that there is one helper (base station) in the network with infinity storage capacity and not making any request, and users who have limited storage capacity make requests (same in our case) but cannot be helpers, by using a sub-packetization based caching and a coded multicasting scheme, the minimum per user throughput scales as $\Theta\left (\max \left \{\frac{1}{n}, \frac{1}{m} \right \} \right )$ and this scheme can achieve a zero outage probability. Interestingly, it has the same order as the minimum per user throughput with an (arbitrarily small) constant outage probability by using our scheme, where the $\Theta\left(\frac{1}{n}\right)$ term in our scheme can be achieved by dividing the network into a constant number of clusters and serving users one by one in each cluster. When $n \gg m$, clearly, our scheme has a large gain comparing to the naive broadcasting scheme but has the same order with the coded multicasting scheme. In order to determine which scheme yields the best performance we have to consider the actual rates for realistic channel physical models 
and not just the scaling laws. This is the object of current investigation. 
However, from a practical implementation viewpoint, we notice that our D2D scheme has very simple caching 
(at random) and delivery phase (one-hop D2D from neighbors).
In contrast, the coded multicasing scheme of \cite{maddah2012fundamental} constructs the cache contents and the coded delivery phase in a combinatorial manner that does not scale well with $n$. For example, in our network configuration, it requires the code length larger than ${10000 \choose 30} \gg 10^{15} $.

\bibliographystyle{IEEEbib}
\bibliography{references}

\end{document}

%% file: macros.tex
\long\def\comment#1{}

\newfont{\bbb}{msbm10 scaled 700}

\newfont{\bb}{msbm10 scaled 1100}

\newcommand{\PP}{\mbox{\bb P}}

\newcommand{\EE}{\mbox{\bb E}}

\newcommand{\Ac}{{\cal A}}

\newcommand{\Ec}{{\cal E}}
\newcommand{\Fc}{{\cal F}}
\newcommand{\Gc}{{\cal G}}
\newcommand{\Hc}{{\cal H}}

\newcommand{\Lc}{{\cal L}}

\newcommand{\Tc}{{\cal T}}
\newcommand{\Uc}{{\cal U}}

\newcommand{\fsf}{{\sf f}}

\newcommand{\Asf}{{\sf A}}

\newcommand{\Gsf}{{\sf G}}

\newcommand{\be}{\begin{equation}}
\newcommand{\ee}{\end{equation}}
\newcommand{\bea}{\begin{eqnarray}}
\newcommand{\eea}{\end{eqnarray}}

\def\fsf{ {\sf f}}